\documentclass[twocolumn,floatfix,superscriptaddress,showpacs,preprintnumbers,prl]{revtex4}
\usepackage{graphicx}
\usepackage{epsfig}
\usepackage{amssymb}

\begin{document}
\title{Identification of network modules by optimization of ratio association}
\author{   L. Angelini$^{\diamondsuit,\heartsuit,\star}$,
           S. Boccaletti$^{\dag}$,
           D. Marinazzo$^{\heartsuit}$,
           M. Pellicoro$^{\diamondsuit,\heartsuit,\star}$ and
           S. Stramaglia$^{\diamondsuit,\heartsuit,\star}$
             \\
\small{$\diamondsuit$ TIRES-Center of Innovative Technologies for
Signal Detection and Processing,
University of Bari, Italy} \\
\small{$\heartsuit$ Dipartimento Interateneo di Fisica, Bari, Italy} \\
\small{$\star$ Istituto Nazionale di Fisica Nucleare, Sezione di Bari, Italy} \\
\small{$\dag$ CNR- Istituto dei Sistemi Complessi,
Via Madonna del Piano, 10, 50019 Sesto Fiorentino (FI), Italy }\\
}

\date{\today}
\begin{abstract}

We introduce a novel method for identifying the modular structures of a network based on
the maximization of an objective function: the ratio association. This cost function
arises when the communities detection problem is described in the probabilistic
autoencoder frame. An analogy with kernel k-means methods allows to develop an efficient
optimization algorithm, based on the deterministic annealing scheme. The performance of
the proposed method is shown on a real data set and on simulated networks.

\end{abstract}
 \maketitle

{\bf \noindent The structure of a complex network may be described by identifying the
modules of which it is composed. The concept of module is qualitative: nodes are more
connected within their modules than with the rest of the network, and its quantification
is still a subject of debate. Modularity, a quantity related to the correlation between
the probability of having an edge joining two sites and the fact that the sites belong to
the same modules, has been widely accepted as a measure for module identification. Here
we provide a new description of this important problem. We analyze the use of a novel
objective function, the ratio association, measuring the coherency between modules. Ratio
association emerges in the probabilistic autoencoder frame, performing a lossy
compression of the network's structures. An analogy to kernel k-means allows the
development of an efficient algorithm for the optimization of ratio association. The
power of the proposed technique is assessed by showing the structures found by ratio
optmization on a real data-set and on simulated networks. The likelihood of the
probabilistic autoencoder model may be used to select the optimal number of modules.

}
\section{Introduction} A hierarchical
structure of modules characterizes the topology of most of real-world networked systems
\cite{boccaletti}. In social networks, for instance, these modules are densely connected
groups of individuals belonging to social communities. {\it Modules} (called also {\it
community structures}) are defined as tightly connected subgraphs of a network, i.e.
subsets of nodes {\it within} which the density of links is very high, while {\it
between} which connections are much sparser. These tight-knit modules constitute units
that separately (and in parallel) contribute to the collective functioning of the
network. For instance, the presence of subgroups in {\it biological} and {\it
technological} networks is at the basis of their functioning. Hence the issue of
detecting and characterizing module structures in networks received considerable amount
of attention.

Rigorously, the identification of the hierarchy of modules of a network is equivalent to
the {\it graph partitioning} problem in computer science, which is known to be a
NP-complete problem \cite{gj79}. A series of efficient heuristic methods has been
proposed over the years to cope with this problem. These include methods based on {\it
spectral analysis} \cite{spectral}, or {\it hierarchical clustering methods} developed in
the context of social networks analysis \cite{hc}. Among the different techniques, we
recall the modularity identification based on the statistical properties of a system of
spins \cite{potts}, and hierarchical clustering techniques exploiting the central concept
of modularity \cite{community,gn}. The modularity $Q$ is a measure of the correlation
between the probability of having an edge joining two sites and the fact that the sites
belong to the same modules (see Ref.~\cite{gn} for the mathematical definition of $Q$).
Methods directly based on the optimization of $Q$ have been proposed \cite{amaral,duch},
while recently a spectral technique has been introduced \cite{newman} exploiting the
information of the {\it modularity matrix}, that, for a given graph, has the property of
being dense even when the adjacency matrix is sparse.

Furthermore, another recent stream of research has been initiated
by the relevant observation that topological hierarchies are
associated to dynamical time scales in the transient of a
synchronization process \cite{arenas}. Such an observation
inspired the introduction of a fast technique able to detect and
identify the modules of a complex network from the cluster
de-synchronization scenario of phase oscillators
\cite{boccacommunities}.

In this paper, we introduce a new technique for modules identification, that efficiently
optimizes an objective function called {\it ratio association}. Precisely, once the
number of modules $n_c$ is fixed, the optimization process leads to a fast (in a time
scaling linearly with the number of nodes $N$ in the network) detection of the
corresponding modules.  The efficiency of the algorithm is due to an equivalence with the
kernel k-means \cite{girolami}, which we exploit in the deterministic annealing frame
\cite{da}. We show that the optimization of ratio association may be motivated in the
probabilistic autoencoder frame, a paradigm which has been used to derive cost functions
for data clustering \cite{nitti}; the same cost function has been used in \cite{schr} for
classification of time series data. In order to select the number of modules $n_c$, the
quality of the solution is to be assessed. This can be achieved in two ways, on the basis
of the modularity of the solution, or according to the likelihood of the probabilistic
autoencoder model.

The paper is organized as follows. In the next Section we describe our method, while in
Section 3 some applications are shown. Section 4 summarizes our conclusions.
%%%%%%%%%%%%%%%%%%%%%%%%%%%%%%%%%%%%%%%%%%%%%%%%%%%%%%%%%%%%%%%%%%%%%%%%%%%
%%%%%%%%%  FIG. 1
\begin{figure}
\begin{center}
\epsfig{figure=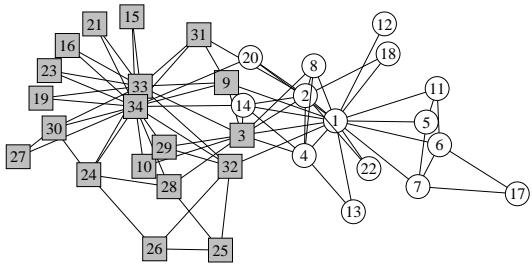,width=7.7truecm,angle=0}
\end{center}
\caption{The Zachary karate club network. The two modules
identified by the proposed algorithm are colored in grey and
white, respectively. Squares and circles indicate the two real
communities described by Zachary \cite{zachari}. Notice that our
technique fully reveals the true subdivision. \label{fig1}}
\end{figure}
%%%%%%%%%%%%%%%%%%%%%%%%%%%%%%%%%%%%%%%%%%%%%%%%%%%%%%%%%%%%%%%
\section{The method}
Given a set of data vectors $\{\mathbf{x}_i\}_{i=1}^N$, with
$\mathbf{x}_i \in {\Bbb R}^n$, the goal of the kernel k-means  is
to find a q-way disjoint partition \cite{notilla}
$\{\pi_c\}_{c=1}^q$ of the data (where $\pi_c$ represents the c-th
cluster) such that the following objective function is minimized:
\begin{eqnarray}
\mathcal{I} \left( \{ \pi_c \}_{c=1}^q \right) = \sum_{c=1}^q
\sum_{\mathbf{x}_i \in \pi_c} || \Phi(\mathbf{x}_i) - \mathbf{m}_c
||^2, \label{obj-kmeans}
\end{eqnarray}
where
\begin{equation}
\mathbf{m}_c ={ \sum_{\mathbf{x}_i \in \pi_c}
\Phi(\mathbf{x}_i)\over |\pi_c|}. \label{proto}
\end{equation}

Here, $|\pi_c|$ is the cardinality of the subset $\pi_c$, and
$\Phi$ is a function mapping the $x$ vectors onto a (generally)
higher-dimensional space (if $\Phi$ is the identity function, the
above Equations recover the standard definition of k-means).

Expanding the distance term $|| \Phi(\mathbf{x}_i) - \mathbf{m}_c
||^2$ in the objective function, one obtains
\begin{eqnarray}
\label{expand} \Phi(\mathbf{x}_i) \cdot \Phi(\mathbf{x}_i)
-{2\sum_{\mathbf{x}_j \in \pi_c} \Phi(\mathbf{x}_i)\cdot
\Phi(\mathbf{x}_j)\over |\pi_c|} \nonumber \\
+ {\sum_{\mathbf{x}_j \in \pi_c} \sum_{\mathbf{x}_\ell \in \pi_c}
\Phi(\mathbf{x}_j)\cdot \Phi(\mathbf{x}_\ell)\over |\pi_c|^2}.
\end{eqnarray}

Notice that in Eq. (\ref{expand}), all computations involving data
points are in the form of inner products. As a result, one can use
the {\it kernel trick}: if one can compute the dot product
$K_{ij}=\Phi(\mathbf{x}_i)\cdot \Phi(\mathbf{x}_j)$ efficiently,
then one is able to compute distances between points in this
mapped space without having to explicitly know the mapping of
$\mathbf{x}_i$ and $\mathbf{x}_j$ onto $\Phi(\mathbf{x}_i)$ and
$\Phi(\mathbf{x}_j)$, respectively. It is known that {\it any}
positive semi-definite matrix $K$ can be thought of as a kernel
matrix \cite{shawe}. Using the kernel matrix, Eq.
(\ref{obj-kmeans}) can be rewritten as:
\begin{eqnarray}
\mathcal{I} \left( \{ \pi_c \}_{c=1}^q \right) = \sum_{c=1}^q
\sum_{i \in \pi_c} (\label{rewritten}K_{ii} -{2\sum_{j \in \pi_c}
K_{ij}\over |\pi_c|} \nonumber \\ + {\sum_{j \in \pi_c} \sum_{\ell
\in \pi_c} K_{j\ell}\over |\pi_c|^2}). \label{cost}
\end{eqnarray}

%%%%%%%%%%%%%%%%%%%%%%%%%%%%%%%%%%%%%%%%%%%%%%%%%%%%%%%%%%%%%%%%%%%%%%%%%%%
%%%%%%%%%  FIG. 2
\begin{figure}
\begin{center}
\epsfig{figure=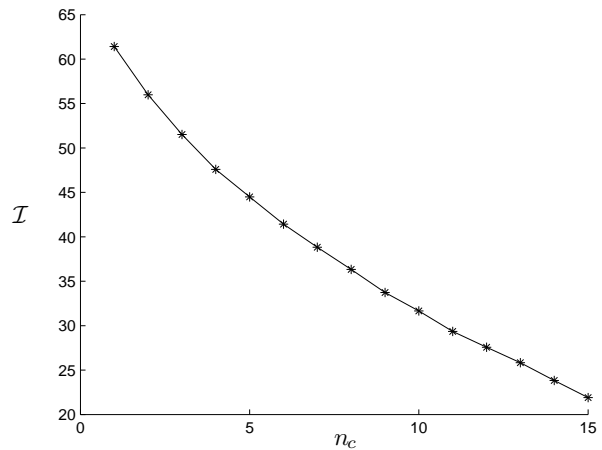,width=8.7truecm,angle=0}
\end{center}
\caption{The objective function $\mathcal{I}$ for the karate club
modular structure found by the proposed algorithm, {\it vs.} the
number of communities $n_C$. Here $\sigma =2$, but  results are
stable against variations of $\sigma$. \label{fig2}}
\end{figure}
%%%%%%%%%%%%%%%%%%%%%%%%%%%%%%%%%%%%%%%%%%%%%%%%%%%%%%%%%%%%%%%

Suppose that the graph $G=(V,A)$ is given, where $V$ is the set of $N$ vertices and $A$
is the adjacency matrix [the elements $A_{ij}$ are one (zero) whenever an edge is (is
not) present between vertices $i$ and $j$]. If $\mathcal{A}$ and $\mathcal{B}$ are two
disjoint subsets of $V$, we furthermore define
$\mbox{links}(\mathcal{A},\mathcal{B})=\sum_{i\in \mathcal{A},j\in \mathcal{B}}A_{ij}$.

The idea is to fix the number of modules $n_c$ into which we want
to efficiently partition the original graph, and to look for the
$n_c$-way disjoint partition of $V$ ($\{\pi_c\}_{c=1}^{n_c}$),
that maximizes the following objective function, called {\it ratio
association} \cite{shi}:
\begin{eqnarray}
\label{obj-ratioass} \mathcal{R} \left( \{ \pi_c \}_{c=1}^{n_c} \right) =
\sum_{c=1}^{n_c} {\mbox{links}(\pi_c,\pi_c)\over |\pi_c|}.
\end{eqnarray}

Let us now associate to the given graph a $N\times N$ kernel
matrix as follows:
\begin{eqnarray}
\label{choice}K= \sigma I + A,
\end{eqnarray}

%%%%%%%%%%%%%%%%%%%%%%%%%%%%%%%%%%%%%%%%%%%%%%%%%%%%%%%%%%%%%%%%%%%%%%%%%%%
%%%%%%%%%  FIG. 3
\begin{figure}
\begin{center}
\epsfig{figure=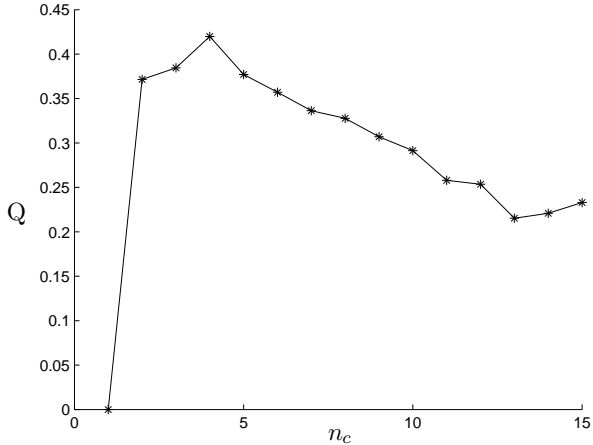,width=8.7truecm,angle=0}
\end{center}
\caption{The modularity $Q$ for the karate club modular structure found by the proposed
algorithm, {\it vs.} $n_c$. As in Fig. \ref{fig2}, we use here $\sigma =2$, but results
are stable against variations of $\sigma$.\label{fig3}}
\end{figure}
%%%%%%%%%%%%%%%%%%%%%%%%%%%%%%%%%%%%%%%%%%%%%%%%%%%%%%%%%%%%%%%

where $I$ is the identity matrix, and $\sigma$ is a real number
chosen to be sufficiently large so that $K$ comes out to be
positive definite. Now, given a $n_c$-way disjoint partition
$\{\pi_c\}_{c=1}^{n_c}$ of the graph, the corresponding value of
the ratio association and the objective function of kernel k-means
are related as follows:
\begin{eqnarray}
\mathcal{I} \left( \{ \pi_c \}_{c=1}^{n_c} \right) =
\left(N-{n_c}\right) \sigma - \mathcal{R} \left( \{ \pi_c
\}_{c=1}^{n_c} \right). \label{relazione}
\end{eqnarray}

An important point follows: $\mathcal{I}$ attains its minimum in correspondence of the
same partition providing the maximum of $\mathcal{R}$, independently of $\sigma$, as it
was shown in Ref. \cite{dhillon} when considering the standard iterations of k-means.
Therefore, the kernel k-means minimization may be straightforwardly used to find the
$n_c$ optimal clustering of the graph, by simply maximizing the ratio association. The
ratio association may be derived in the probabilistic autoencoder frame as described in
the appendix.

%%%%%%%%%%%%%%%%%%%%%%%%%%%%%%%%%%%%%%%%%%%%%%%%%%%%%%%%%%%%%%%%%%%%%%%%%%%
%%%%%%%%%  FIG. 4
\begin{figure}
\begin{center}
\epsfig{figure=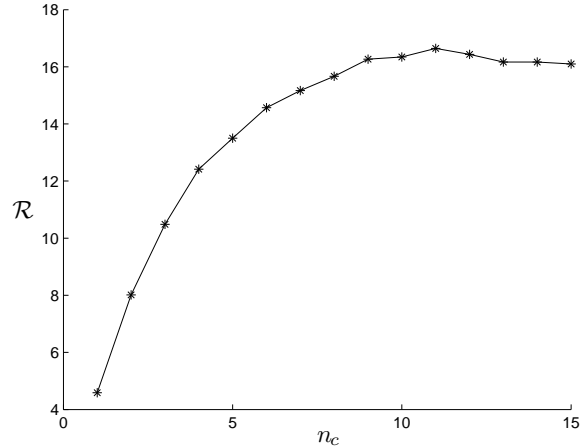,width=8.7truecm,angle=0}
\end{center}
\caption{The ratio association $\mathcal{R}$ for the karate club modular structure found
by the proposed algorithm, {\it vs.} $n_C$. Same stipulations on $\sigma$ as in the
captions of Figs. \ref{fig2}, \ref{fig3}. \label{fig4}}
\end{figure}
%%%%%%%%%%%%%%%%%%%%%%%%%%%%%%%%%%%%%%%%%%%%%%%%%%%%%%%%%%%%%%%

Hence, we can use graph clustering to discover modules structures.
As we here deal with modular structures maximizing the ratio
association, in the following we will handle the optimization
problem by deterministic annealing.

Let $\rho_{ic}$ be the probability that vertex $i$ belongs to the
$c$-th module. We write
\begin{equation}
\label{prob}\rho_{ic}= {e^{-\beta \xi_{ic}}\over \sum_{c'=1}^{n_c}
e^{-\beta \xi_{ic'}}},
\end{equation}
where, according to (\ref{cost}),
\begin{equation}
\label{fields}\xi_{ic}= K_{ii} -{2\sum_{j=1}^N K_{ij}\;\rho_{jc}\over \sum_{m=1}^N
\rho_{mc}} + {\sum_{j, \ell=1}^N K_{j\ell}\;\rho_{jc}\rho_{\ell c}\over
\left(\sum_{m=1}^N \rho_{mc}\right)^2},
\end{equation}
and $K$ is given by (\ref{choice}).

Starting from a random configuration of $\{\rho \}$ and $\{\xi
\}$, Eqs. (\ref{prob}-\ref{fields}) are solved iteratively while
exponentially increasing $\beta$. At large $\beta$, $\rho_i(c)$
are all zero except for one element providing the module to whom
the vertex $i$ has to be assigned.

Notice that the annealing procedure leads to a final partition of vertices which still
has a tiny dependance on the starting configuration, hence the algorithm is to be run
several times, selecting the partition leading to the lowest value of $\mathcal{I}$. As
is typical of deterministic annealing approaches, the complexity of the algorithm is
$\mathcal{O}\left( n_c N \langle z\rangle\right)$, where $\langle z\rangle$ is the
average number of edges per vertex. Note, however, that in the proposed method the number
of modules is to be specified in advance. Therefore, for a full hierarchical description
of the original network (i.e. when one wants also to determine the optimal $n_c$), the
algorithm has to be run by varying $n_c$ between its minimum (2) and its maximum ($N$)
value, leading to an overall complexity $\mathcal{O}\left( N^3 \langle z\rangle\right)$.

%%%%%%%%%%%%%%%%%%%%%%%%%%%%%%%%%%%%%%%%%%%%%%%%%%%%%%%%%%%%%%%%%%%%%%%%%%%
%%%%%%%%%  FIG. 5
\begin{figure}
\begin{center}
\epsfig{figure=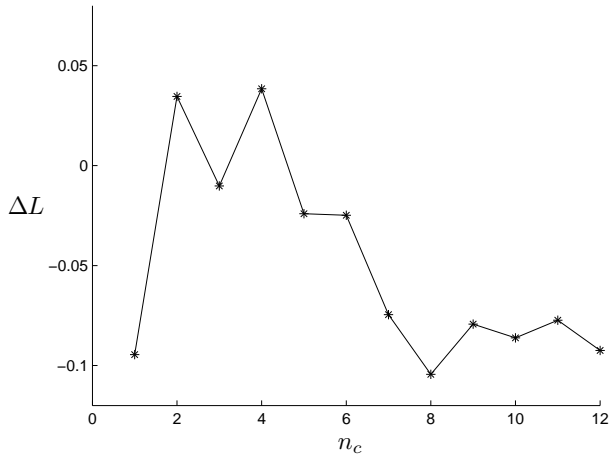,width=8.7truecm,angle=0}
\end{center}
\caption{$\Delta L=L-L_0$ (see the text) is plotted versus $n_c$, for the Zachary
network. \label{deltak}}
\end{figure}
%%%%%%%%%%%%%%%%%%%%%%%%%%%%%%%%%%%%%%%%%%%%%%%%%%%%%%%%%%%%%%%

The choice of $\sigma$ deserves few comments. As said before, to
enforce the positive definiteness of $K$, and thus to establish
the connection to kernel k-means, $\sigma$ must be sufficiently
large. However, since varying $\sigma$ does not change the global
optimum, one may choose $\sigma$ in the most convenient way from
the computational point of view, even though $K$ will not be
ensured to be positive definite.

\section{Applications}

Let us first discuss the application of the proposed method to the well-known Zachary
karate club network \cite{zachari}, shown in Figure \ref{fig1}. We point out that the
output of the algorithm is independent on $\sigma \in [0, 20]$ (notice that $K$, in this
case, is positive for $\sigma
> 5$).

When selecting $n_c=2$ (i.e. when trying to split the network in
two modules), Figure \ref{fig1} shows that one fully recovers the
true subdivision of the data set, with $\mathcal{R}= 8.0139$ and
modularity $Q= 0.3715$.

When repeating the analysis  at varying $n_c$, Figure \ref{fig2} reports the value of
$\mathcal{I}$, corresponding to the solution, as a function of $n_c$: it is a strictly
decreasing function. In Figure \ref{fig3} we plot the modularity $Q$ of the solution
versus $n_C$: the maximum is $Q=0.420$ and corresponds to a partition of the graph into
four modules, in perfect agreement with the outcome of other techniques previously tested
on the Zachary karate club network. Finally, Figure \ref{fig4} reports the ratio
association $\mathcal{R}$ versus $n_C$, making it evident the validity of Eq.
(\ref{relazione}).

The selection of $n_c$ may also be done on the basis of the average log-likelihood of the
autoencoder (see the appendix). In Figure \ref{deltak} we plot $L-L_0$ vs $n_c$, where
$L$ is the average log-likelihood of the data set, whilst $L_0$ is the same quantity
evaluated on a network with the same number of nodes and  links but with links randomly
assigned to pairs of nodes \cite{ellezero}. According to the criterion of the largest
$\Delta L = L-L_0$, both $n_c =2$ and $n_c =4$ are suitable partitions.

%%%%%%%%%%%%%%%%%%%%%%%%%%%%%%%%%%%%%%%%%%%%%%%%%%%%%%%%%%%%%%%%%%%%%%%%%%%
%%%%%%%%%  FIG. 6
\begin{figure}
\begin{center}
\epsfig{figure=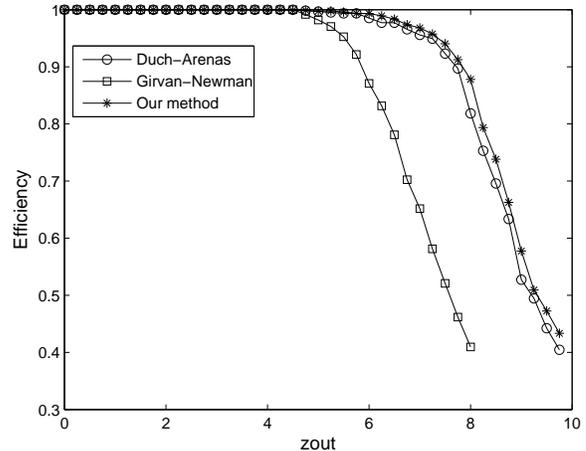,width=8.7truecm,angle=0}
\end{center}
\caption{The fraction $p$ of correctly classified nodes is plotted versus $z_{out}$, the
average number of edges a node forms with members of other modules, for the proposed
algorithm (stars), for the Girvan and Newman method \cite{gn} (empty squares) and for the
method introduced by Duch and Arenas \cite{duch} (empty circles). Each point refers to an
ensemble average over 100 different network realizations. Same stipulations for $\sigma
=2$ as in the Captions of Figs. \ref{fig2}-\ref{fig4}. \label{fig5}}
\end{figure}
%%%%%%%%%%%%%%%%%%%%%%%%%%%%%%%%%%%%%%%%%%%%%%%%%%%%%%%%%%%%%%%
%%%%%%%%%%%%%%%%%%%%%%%%%%%%%%%%%%%%%%%%%%%%%%%%%%%%%%%%%%%%%%%%%%%%%%%%%%%
%%%%%%%%%  FIG. 7
\begin{figure}
\begin{center}
\epsfig{figure=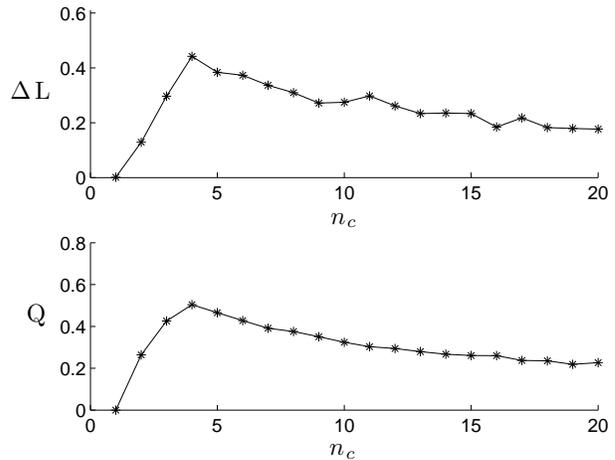,width=8.7truecm,angle=0}
\end{center}
\caption{Top: $\Delta L=L-L_0$ (see the text) is plotted versus $n_c$, for a randomly
generated network with $z_{out}=4$. Bottom: the modularity of the modular structure found
by the proposed algorithm on the same network.\label{deltanet}}
\end{figure}
%%%%%%%%%%%%%%%%%%%%%%%%%%%%%%%%%%%%%%%%%%%%%%%%%%%%%%%%%%%%%%%
%%%%%%%%%%%%%%%%%%%%%%%%%%%%%%%%%%%%%%%%%%%%%%%%%%%%%%%%%%%%%%%%%%%%%%%%%%%
%%%%%%%%%  FIG. 8
\begin{figure}
\begin{center}
\epsfig{figure=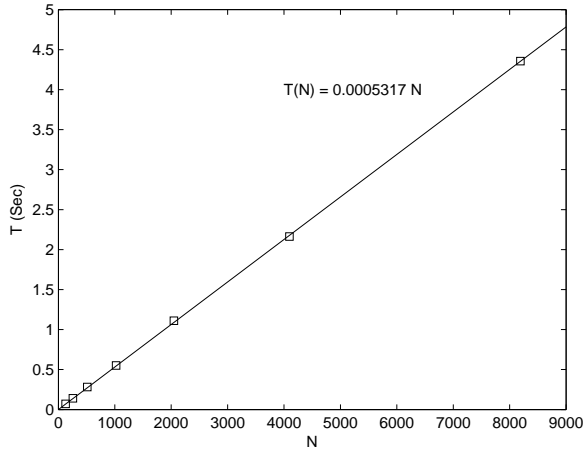,width=8.7truecm,angle=0}
\end{center}
\caption{The  scaling of the CPU time is reported as a function of the number of vertices
$N$, at fixed $n_c$.\label{fig8}}
\end{figure}
%%%%%%%%%%%%%%%%%%%%%%%%%%%%%%%%%%%%%%%%%%%%%%%%%%%%%%%%%%%%%%%

To evaluate the performance of the proposed technique, we generate a set of random graphs
featuring a well defined modular structure. Precisely, all graphs are generated with $N =
128$ nodes and $K=1024$ edges. The nodes are distributed into four modules, containing 32
nodes each. Pairs of nodes belonging to the same module (to different modules) are linked
with probability $p_{in}$ ($p_{out}$). $p_{out}$ is taken so that the average number
$z_{out}$ of edges a node forms with members of other communities can be controlled (in
our trials $z_{out}$ has been varied between $0$ and $10$). $p_{in}$ is chosen so as to
maintain a constant total average node degree $<k>=16$. Notice that, as $z_{out}$
increases, the modular structure of the network becomes weaker and harder to identify. As
the {\it real} modular structure is here directly imposed by the generation process, the
performance of the identification method can be assessed by monitoring the fraction $p$
of correctly classified nodes vs. $z_{out}$. In Figure \ref{fig5}, we report a
comparative analysis of $p$ {\it vs.} $z_{out}$ for the proposed algorithm, for the
Girvan and Newman method \cite{gn} and for the method introduced by Duch and Arenas
\cite{duch}. The result is that the accuracy attained by our method comes out to be
slightly better than that of Ref. \cite{duch}. Note that while applying our algorithm to
these networks, we have selected $n_c$ by maximizing $L-L_0$ (similar results are
obtained maximizing the modularity, see Figure \ref{deltanet}).

We finally report in Figure \ref{fig8} the CPU time needed to complete a given partition
as a function of the number of vertices. The curve confirms that, for a given $n_c$, the
computational demand scales linearly with the network size.

\section{Conclusions}

In conclusion, we introduced a novel method for identifying the modular structures of a
network based on the maximization of an objective function: the ratio association. This
objective function emerges in the frame of probabilistic autoencoders, thus providing a
new description of the communities detection problem, in terms of a lossy compression of
the structures, as well as a new selection strategy for the number of modules by means of
the log-likelihood of the model.
\section{Acknowledgements}
The Authors are indebted with V. Latora, Y. Moreno and G. Nardulli for the many helpful
discussions on the subject. S.B. acknowledges the Yeshaya Horowitz Association through
the Center for Complexity Science.

\section{Appendix}
In this appendix we show that ratio association may be derived in the Probabilistic
Autoencoder Framework. We briefly discuss autoencoders described by one-stage folded
Markov chains \cite{nitti}. Let us consider a point $x$, in a data space, sampled with
probability distribution $P_0 \left(x\right)$; a code index $\alpha \in \{1,\ldots,q\}$
is assigned to $x$ according to conditional probabilities $P\left( \alpha |x\right)$. A
reconstructed version of the input, $x'$, is then obtained by use of the Bayesian
decoder:
\begin{equation}
P\left( x'|\alpha\right)={P\left( \alpha |x'\right) P_0 \left(x'\right)\over P\left(
\alpha\right)}. \label{eq:bayes}
\end{equation}
The joint distribution of $x$, $x'$ and $\alpha$, describing this encoding-decoding
process, is
\begin{equation}
P\left( x,x',\alpha\right)=P_0 \left( x\right)P\left( \alpha |x\right)
P\left(x'|\alpha\right); \label{eq:bayes1}
\end{equation}
owing to (\ref{eq:bayes}), the joint distribution reads:
\begin{equation}
P\left( x,x',\alpha\right)={P_0 \left(x\right)P_0 \left(x'\right) P\left( \alpha
|x\right) P\left( \alpha |x'\right) \over P\left( \alpha\right)}. \label{eq:bayes2}
\end{equation}
The conditional probabilities $\{ P\left( \alpha |x\right)\}$ are the free parameters
that must be adjusted to force the autoencoder to emulate the identity map on the data
space.

Let s(x,x') be a measure of the similarity between input and output; the average
similarity is then given by
\begin{equation}
{\cal S}=\sum_{\gamma=1}^q \int dx \int dx'{P_0 \left(x\right)P_0 \left(x'\right) P\left(
\alpha |x\right) P\left( \alpha |x'\right) \over P\left( \alpha\right)} s(x,x').
\label{eq:aves}
\end{equation}
A good autoencoder is obviously characterized by a high value of ${\cal S}$. Given a set
of data vectors $\{\mathbf{x}_i\}_{i=1}^N$, partitioning these points in $q$ modules
corresponds, in this frame, to design an autoencoder, with $q$ code indexes, acting on
data space. Choosing the encoder to be deterministic leads to:
\begin{equation}
P\left( \alpha |x\right)=\delta_{\alpha\;\gamma(x)}, \label{eq:delta}
\end{equation}
$\gamma (x)\in \{1,\ldots,q\}$ being the code index associated to $x$. The estimate for
the average similarity, based on the data-set at hand, is given by :
\begin{equation}
\hat{{\cal S}}= {1\over N}\sum_{\alpha=1}^q {\sum_{i,j=1}^N \delta_{\alpha
\gamma_i}\delta_{\alpha \gamma_j}s_{ij}\over \sum_{k=1}^N \delta_{\alpha \gamma_k}}.
\label{hams}
\end{equation}
If the similarity matrix $s_{ij}$ is identified with the kernel matrix $K$, we obtain:
$$N \hat{{\cal S}}={n_c} \sigma + \mathcal{R}.$$
Therefore maximization of the ratio association is equivalent to design the most
effective autoencoder, the effectiveness being measured by the average similarity.

Now we consider the average log-likelihood \cite{papoulis} of data
$\{\mathbf{x}_i\}_{i=1}^N$:
\begin{equation}
\sum_{\gamma=1}^q \int dx P\left( x,\gamma\right) \log{ P\left( x |\gamma\right)}.
\label{eq:prob}
\end{equation}
We may easily obtain an estimate of this quantity, which measures how good is the
autoencoder frame to model the data-set. Using kernel density estimation \cite{duda} we
easily obtain:
\begin{equation}
L={1\over N} \sum_{i=1}^N \log\left( {\mbox{links}\left( \{i\},\pi_{\gamma(i)} \right)
\over |\pi_{\gamma(i)}|}\right); \label{eq:likeli}
\end{equation}
the numerator, in the formula above, is the number of links from node $i$ to nodes in the
same module as $i$, whereas the denominator is the cardinality of the module whom $i$
belongs to.


\begin{thebibliography}{99}
\bibitem{notilla} By q-way disjoint partition it is meant
a set of $q$ disjoint data subsets, whose union is the whole data.
\bibitem{ellezero} We randomly generate $100$ such random networks and take $L_0$ as the average;
however we find that $L_0$ varies less than $1\%$ over the realizations.
\end{thebibliography}
\end{document}